\documentclass[aps,prd,showpacs,showkeys,twocolumn]{revtex4}
\usepackage{epsfig,amsmath,amsfonts,amssymb,graphics,rotating}
\begin{document}
\title{\bf Non-minimally coupled multi-scalar black holes}

\author{Ben M. Leith}
\email{b.leith@phys.canterbury.ac.nz}
\affiliation{Department of
Physics and Astronomy, University of Canterbury, Private Bag 4800,
Christchurch, New Zealand}
\author{Alex B. Nielsen}
\email{abn16@student.canterbury.ac.nz}
\affiliation{Department of
Physics and Astronomy, University of Canterbury, Private Bag 4800,
Christchurch, New Zealand}
\affiliation{Center for Theoretical
Physics and School of Physics, College of Natural Sciences, Seoul
National University, Seoul 151-742, Korea}
\begin{abstract}
We study the static, spherically symmetric black hole solutions
for a non-minimally coupled multi-scalar theory. We find numerical
solutions for values of the scalar fields when a certain
constraint on the maximal charge is satisfied. Beyond this
constraint no black hole solutions exist. This constraint
therefore corresponds to extremal solutions, however, this does
not match the $\kappa = 0$ constraint which typically indicates
extremal solutions in other models. This implies that the set of
extremal solutions have non-zero, finite and varying surface
gravity. These solutions also violate the no-hair theorems for
$N>1$ scalar fields and have previously been proven to be linearly
stable.

\centerline{arxiv:0709.2541}

\medskip

\centerline{17 September 2007; \LaTeX-ed \today }
\end{abstract}
\maketitle
\newcommand{\beq}{\begin{equation}}
\newcommand{\eeq}{\end{equation}}
\newcommand{\bea}{\begin{eqnarray}}
\newcommand{\eea}{\end{eqnarray}}
\def\d{{\mathrm{d}}}
\def\Tr{{\mathrm{Tr}}}
\section{Introduction}

Black hole solutions with scalar fields are usually constrained to
possessing only {\it secondary hair} by the ``no-hair
conjectures''~\cite{Bekenstein:1971hc}. Early attempts to find
black hole solutions coupled solely to a scalar field found
solutions where the scalar field diverged at the putative
horizon~\cite{Bekenstein:1974sf}. Hence, technically such
solutions cannot actually be considered black hole solutions due
to a lack of a regular horizon. They are unlikely to have any
physical relevance~\cite{Sudarsky:1997te}. The original ``no-hair
conjectures'' have since been violated in a number of cases,
either via coupling the scalar fields to both gravity and gauge
fields, or through violation of the dominant energy condition.
When the existence of scalar hair depends on a non-vanishing gauge
field, and is entirely fixed by the mass, gauge charge and angular
momentum this is called secondary hair~\cite{Coleman:1991jf}. In
this paper we discuss a solution with contingent primary
hair~\cite{Mignemi:2004ms}, that is to say the scalar hair depends
on the existence of a non-vanishing gauge field but its behaviour
is not entirely fixed by the values of the other asymptotic
parameters. We briefly list some further ``hairy'' black hole
solutions.

\begin{itemize}

\item{Scalar fields coupled to higher order gravity have been
heavily investigated since they arise naturally in low energy
effective 4-dimensional string theories. The Gauss-Bonnet term,
which is the only ghost-free leading order curvature correction,
has naturally been of particular
interest~\cite{Mignemi:1992nt,Chen:2006ge}.}

\item{Minimally coupled scalar fields with dominant energy
condition violating potentials have been shown to allow
non-trivial hair~\cite{Bechmann:1995sa, Dennhardt:1996cz,
Bronnikov:2001ah, Nucamendi:1995ex}. Examples have been found both
analytically and numerically provided there is at least one global
minimum with $V(\phi) < 0$.}

\item{Theories which couple gravity to non-Abelian gauge fields
such as Einstein-Yang-Mills, Einstein-Yang-Mills-Higgs and
Einstein-Skyrme, usually contain nonlinear self-interactions and
admit ``hairy'' black holes. Einstein-Yang-Mills-Higgs and
Einstein-Skyrme also include scalar fields. These vanish
exponentially at infinity, however, and thus they do not have
``Gauss-like'` scalar charge. These hairy black holes were thought
to be generally unstable but it has been shown that some branches
of solutions of the Einstein-Skyrme black holes are linearly
stable~\cite{Heusler:1992av}. Whether they are non-linearly stable
remains an open question.}

\item{Scalar fields non-minimally coupled to an Abelian gauge
theory have been shown to have hairy
solutions~\cite{Dobiasch:1981vh,Gibbons:1985ac,Gibbons:1987ps,Garfinkle:1990qj}.
Such theories arise naturally in Kaluza-Klein theories and
effective low-energy limits of string theory with a non-trivial
dilaton.}

\end{itemize}

Despite the solutions listed above being beyond the premises of
the original ``no-hair conjecture'', they are still considered
interesting as tests of the limits of the conjectures. Stability
is still an open problem in most cases.

For a non-rotating, static black hole with a single scalar field
coupled to the $U(1)$ electromagnetic gauge
field~\cite{Gibbons:1987ps}. This solution is not a member of the
Reissner--Nordstr\"om class but is entirely specified by the
values of M, Q and P. Adding an extra scalar field was shown to
give more freedom~\cite{Mignemi:1999zy} and a version of scalar
hair that falls between the definitions of primary and secondary
hair. This was called {\it contingent primary hair} and has been
generalised to $N$ scalar fields with linear stability being
shown~\cite{Mignemi:2004ms}. Here we present numerical solutions
to this model and discuss some of the features. As we will see,
the most interesting feature is that these solutions limit to
non-zero, finite surface gravity for an extremal black hole
solution with a general coupling.

This paper is organised as follows. Section \ref{sec:model}
discusses the model to be used and the analytical constraints that
can be placed on the solutions. Section \ref{sec:numerics}
contains the main results and gives details of the solutions found
while the thermodynamic behaviour of these solutions is shown in
section \ref{sec:thermodynamics}. We conclude with a discussion in
section \ref{sec:discussion}. We use the notation of
\cite{Mignemi:2004ms} and define $c=4\pi\kappa^{2}=1$.

\section{Model}
\label{sec:model}

The general Lagrangian density for the $N$-scalar field case is
\bea
{\cal{L}} = \frac{1}{4}\Bigg[ R &-& 2 \Lambda -
2\sum^{N}_{i=1}\partial^{\mu}\Phi_{i}\partial_{\mu}\Phi_{i}
\nonumber \\ &-& \left(
{\sum^{N}_{i=1}\lambda_{i}^{2}}\right)^{-1}\sum^{N}_{i=1}\lambda_{i}^{2}e^{-2g_{i}\Phi_{i}}
F_{\mu\nu}F^{\mu\nu}\Bigg],\label{genlag}
\eea
where $R$ is the Ricci scalar and $F_{\mu\nu}$ is the
electromagnetic field strength. Initially we consider the $N = 2$
case with no cosmological constant, ie. $\Lambda = 0$. For
simplicity we split the representation of the scalar fields such
that
\bea
{\cal{L}} = \frac{1}{4}\Bigg[ R &-&
2\partial^{\mu}\Phi\partial_{\mu}\Phi - 2
\partial^{\mu}\Psi\partial_{\mu}\Psi \nonumber
\\ &-& \frac{\lambda_{1}^{2} e^{-2 g_{1} \Phi} + \lambda_{2}^{2} e^{-2 g_{2}
\Psi}}{\lambda_{1}^2 + \lambda_{2}^2} F_{\mu\nu}F^{\mu\nu}\Bigg],
\label{2scllag}
\eea
Since there is no potential dependent on any of the scalar fields,
the Lagrangian density has the same scale invariance as the
Gibbons-Maeda solution~\cite{Gibbons:1987ps}. This invariance
applies under global re-scalings of the metric $g_{ab} \rightarrow
\omega^{2}g_{ab}$ where $\nabla_{a}\omega = 0$.

We use a standard metric ansatz for static, spherically symmetric
Schwarzschild coordinates following the formalism of
\cite{Nielsen:2005af}.
\bea \label{metric} \d s^{2} = - A^{2}(r)\left(1 -
\frac{2m(r)}{r}\right)\d t^{2} \nonumber \\ +
\frac{1}{1-\frac{2m(r)}{r}}\d r^{2} + r^{2}\d\Omega^{2} \eea
where $\d\Omega^{2} = \d\theta^{2} + \sin^{2}\theta\d\phi^{2}$ and
$m(r)$ is the familiar Misner-Sharp mass function. In order for
non-trivial solutions to exist we take the magnetic monopole field
ansatz
\beq F_{\theta\phi} = P\sin\theta \eeq
where $P$ is the magnetic charge. This choice is made out of
convenience. Due to the scalar coupling to the electromagnetic
sector, the electric ansatz includes dependence on the scalar
fields and is therefore non-trivial, the magnetic ansatz, being
the $\theta \phi$ components of the electromagnetic tensor, avoids
these complications. There is no longer a simple duality between
the magnetic and electric solutions although solutions for the
electric solution should still be tractable if the magnetic
solutions exist. We could, of course, also consider a situation
where both are non-zero. As in the single scalar field of
\cite{Gibbons:1987ps} the scalar fields will necessarily vanish if
$Q=P=0$.

The $G_{t}^{\hspace{0.1cm}t}$ component of the Einstein equations
gives
\bea \label {eomGtt} \frac{2m'}{r^2} = \Bigg(1 &-&
\frac{2m(r)}{r}\Bigg)\left(\Phi'^{2}+\Psi'^{2}\right) \nonumber
\\ &+& \left(\frac{\lambda_1^2 e^{-2g_1\Phi} + \lambda_2^2 e^{-2g_2
\Psi}}{\lambda_1^2 + \lambda_2^2} \right) \frac{P^{2}}{r^{4}}.
\eea
The linear combination
$G_{t}^{\hspace{0.1cm}t}-G_{r}^{\hspace{0.1cm}r}$ of components of
the Einstein equations gives
\beq \frac{A'}{A} = r\left(\Phi'^{2}+\Psi'^{2}\right),
\label{2scalfe2} \eeq
while the two scalar field equations are
\beq
\partial_{r}\left(\left(1-\frac{2m(r)}{r}\right)Ar^{2}\partial_{r}\Phi\right)
= - A\lambda_{1}^{2}g_{1}e^{-2g_{1}\Phi}\frac{P^{2}}{r^{2}}
\label{2scalfe3} \eeq
and
\beq
\partial_{r}\left(\left(1-\frac{2m(r)}{r}\right)Ar^{2}\partial_{r}\Psi\right)
= - A\lambda_{2}^{2}g_{2}e^{-2g_{2}\Psi}\frac{P^{2}}{r^{2}}.
\label{2scalfe4} \eeq
We note these generalise to $N+2$ field equations for the system
given in (\ref{genlag}),
\bea \frac{2m'}{r^2} = \Bigg(1 &-&
\frac{2m(r)}{r}\Bigg)\left(\sum^N_{i=1}\Phi_i'^{2}\right)
\nonumber \\ &+& \left( \sum_{i=1}^N \lambda_i^2 \right)^{-1}
\sum^N_{i=1} \lambda_i^2 e^{-2 g_i \Phi_i} \frac{P^{2}}{r^{4}} +
\Lambda, \eea
\beq \frac{A'}{A} = r\sum^N_{i=1}\Phi_i'^{2} \eeq
and
\beq
\partial_{r}\left(\left(1-\frac{2m(r)}{r}\right)Ar^{2}\partial_{r}\Phi_i\right)
= - A\lambda_{i}^{2}g_{i}e^{-2g_{i}\Phi_i}\frac{P^{2}}{r^{2}}.
\eeq
\section{Numerical Solutions}
\label{sec:numerics}
\begin{figure*}[]
\centering \mbox{\resizebox{8.0cm}{!}{\includegraphics{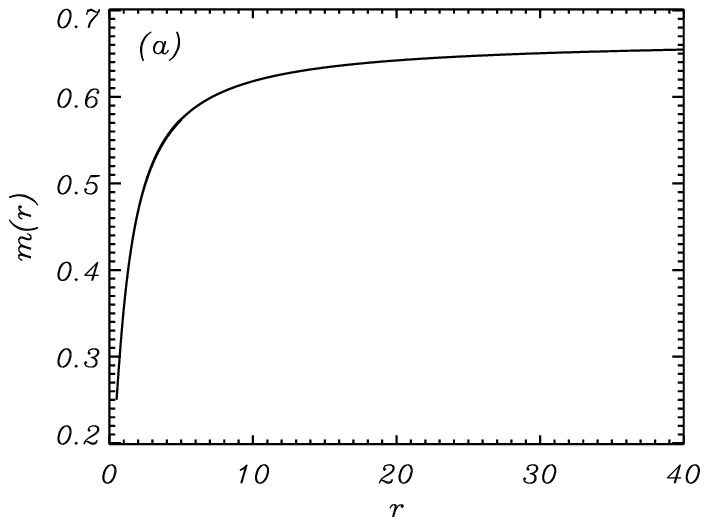}}}
\mbox{\resizebox{8.0cm}{!}{\includegraphics{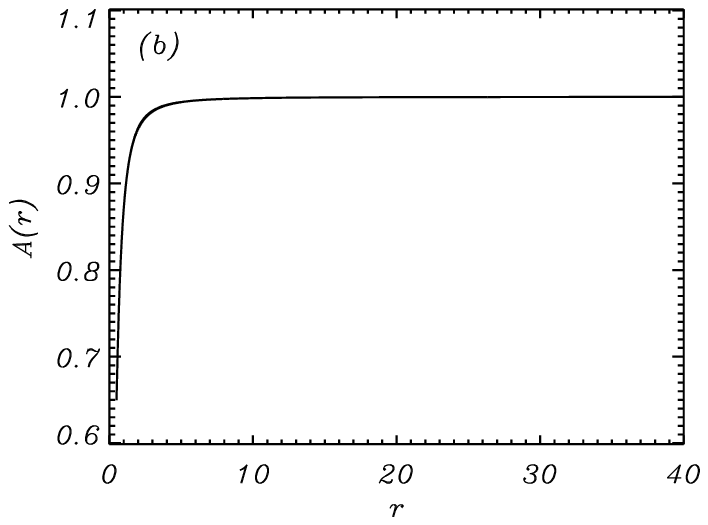}}}
\mbox{\resizebox{8.0cm}{!}{\includegraphics{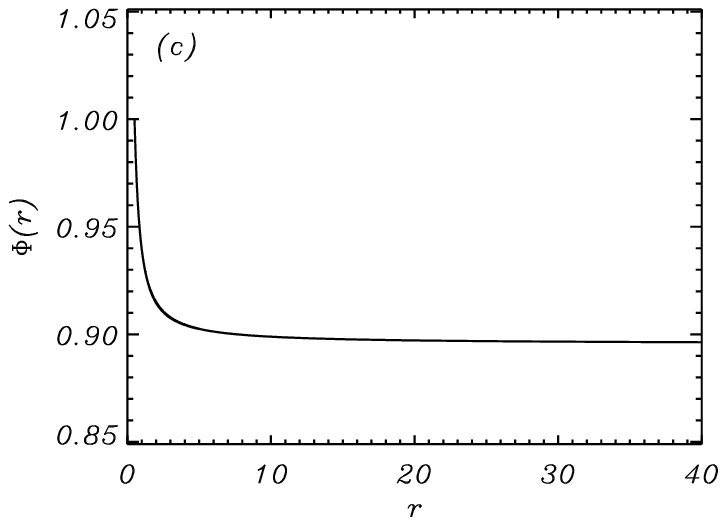}}}
\mbox{\resizebox{8.0cm}{!}{\includegraphics{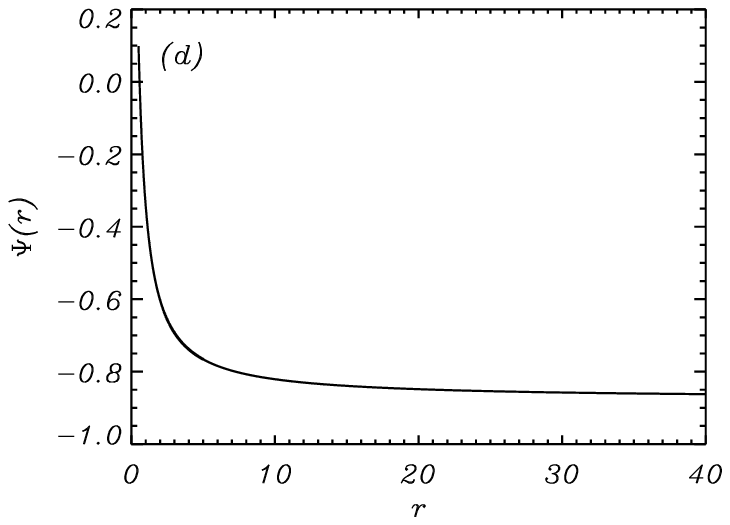}}}
\caption{Solution given by
$\lambda_{1}=\lambda_{2}=g_{1}=g_{2}=1.0$, $P = 0.5$ and with
horizon values $m(r_h)=0.25$, $\Phi(r_h)=1.0$ and $\Psi(r_h)=0.1$.
$P_\infty = 0.8611$ and $M_{ADM} = 0.6674$ while $A(r_h)=0.6490$
as found by the shooting method: (a) $m(r)$; (b) $A(r)$; (c)
$\Phi(r)$; (d) $\Psi(r)$.} \label{figure1}
\end{figure*}
The $N=2$ solutions can be obtained numerically with the help of
the following expansions near the horizon;
\bea m(r) & = & m_{h} + m_{1}(r-r_{h}) + m_{2}(r-r_{h})^2 + ...,
\nonumber \\ A(r) & = & A_{h} + A_{1}(r-r_{h}) + A_{2}(r-r_{h})^2
+ ..., \nonumber \\ \Phi(r) & = & \Phi_{h} + \Phi_{1}(r-r_{h}) +
\Phi_{2}(r-r_{h})^2 + ..., \nonumber \\ \Psi(r) & = & \Psi_{h} +
\Psi_{1}(r-r_{h}) + \Psi_{2}(r-r_{h})^2 + .... \label{expansions}
\eea
In Fig. \ref{figure1} we show the general form of the solutions
for a non-extremal case.

The solution is uniquely defined by 3 asymptotic charges in the
$N=2$ case and by $N+1$ charges in the general case. The Arnowitt,
Deser and Misner (ADM) mass, $M_{ADM}$, is given by the asymptotic
value of $m(r)$, while the asymptotic Gauss-like magnetic charge
is given by
\beq
P_{\infty} = \sqrt{\frac{(\lambda^2_1 e^{-2 g_1 \Phi_{\infty}} +
\lambda_2^2 e^{-2 g_2 \Psi_{\infty}})}{(\lambda_1^2 +
\lambda_2^2)}} P.
\eeq
These two asymptotic charges along with the coefficient
$\Phi_{-1}$ of the $1/r$ term in the asymptotic $\Phi$ expansion
\beq \Phi = \Phi_\infty + \frac{\Phi_{-1}}{r} +
\frac{\Phi_{-2}}{r} + ..., \nonumber \eeq
uniquely define the solution. As shown in \cite{Mignemi:2004ms},
$\Psi_{-1}$ is constrained in the $N=2$ case by
\bea \Phi_{-1}^2 + \Psi_{-1}^2 &+& 2M_{ADM} \left(
\frac{\Phi_{-1}}{g_1} + \frac{\Psi_{-1}}{g_2}\right) \nonumber
\\ &-& \left( \frac{\Phi_{-1}}{g_1} + \frac{\Psi_{-1}}{g_2}
\right)^2 = P_{\infty}^2. \label{cond2} \eea
and in general by
\beq \sum^N_{i=1} \Phi_{-1,\, i}^2 + 2M_{ADM} \sum^N_{i=1}
\frac{\Phi_{-1,\, i}}{g_i} - \left( \sum^N_{i=1} \frac{\Phi_{-1,\,
i}}{g_i} \right)^2 = P_{\infty}^2, \label{condgen}\eeq
where $\Phi_{-1,\, i}$ denotes the $1/r$ coefficient of the $i$th
scalar field. This constraint limits the system to $N+1$ degrees
of freedom, with these being in the magnetic monopole case,
$M_{ADM}$, $P_\infty$ and \{$\Phi_{-1, \, 1}$, $\Phi_{-1, \,
2}$,...,$\Phi_{-1, \, N-1}$\}. The constraint, (\ref{condgen}),
holds throughout our numerical work to the numerical accuracies
required.

We use the shooting method to find solutions such that
\beq \lim_{r \rightarrow \infty} A = 1, \eeq
where this condition is adhered to with a numerical accuracy of
$10^{-7}$. This exploits the rescaling freedom in $A(r_h)$ which
can be seen in (\ref{2scalfe2})-(\ref{2scalfe4}). This is a
necessary requirement for the time coordinate $t$ in
(\ref{metric}) to correspond to the proper time of static
observers at infinity and gives the correct normalisation of the
surface gravity. The numerical limits used to define the
asymptotic region are $m'(r) < 10^{-8}$ and $r
> 500r_h$.
\begin{figure*}[]
\centering \mbox{\resizebox{8.0cm}{!}{\includegraphics{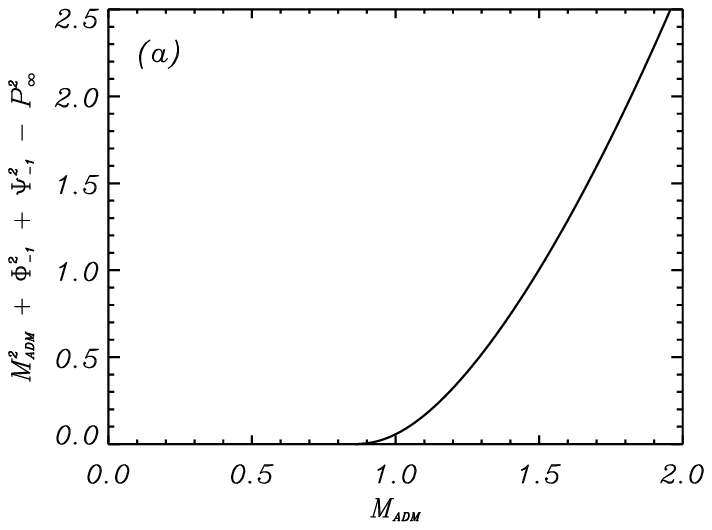}}}
\mbox{\resizebox{8.0cm}{!}{\includegraphics{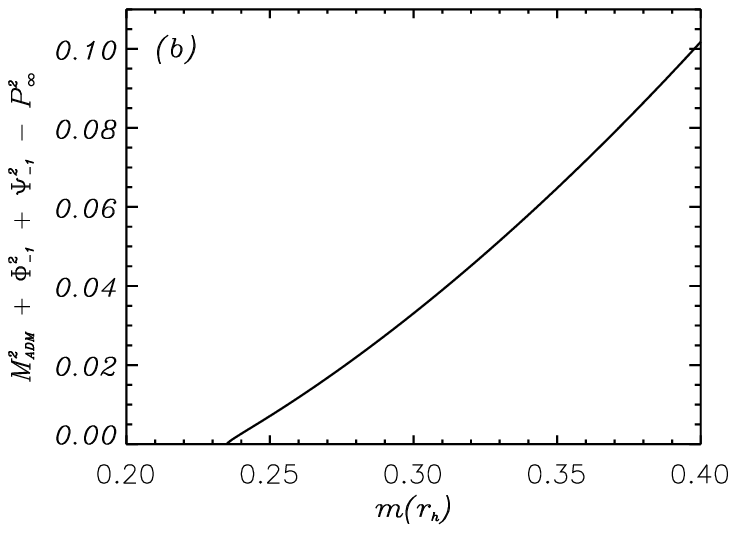}}}
\caption{Behaviour of constraint given in (\ref{cond3}) for
$\lambda_1=\lambda_2=g_1=g_2=1.0$: (a) $P_\infty = 1.2$; shows the
solution approaching the limit as it reaches the extremal case at
$M_{ADM} =0.849$ (numerical accuracy of $10^{-5}$ used for finding
$P_\infty$ and $M_{ADM}$), note that although (a) shows the
expected behaviour it does not allow comparison with the expected
values for $\kappa = 0$ as all quantities are defined at infinity;
(b) $P = 0.5$; approaches the limit $m(r_h) = 0.233$ while the
required value for $\kappa = 0$ from (\ref{cond1}) is
$m(r_h)=0.173$.} \label{figure2}
\end{figure*}

We have also found solutions to both the $N=3$, $\Lambda = 0$ and
$N=2$, $\Lambda = -1$ cases. The $N=3$ case appear
elsewhere~\cite{leiththesis} as there are no particular additional
features when compared to the solutions in fig. \ref{figure1}.
From these results, however, we would assume that solutions, with
$N>2$ scalar fields, exist, having $N+1$ degrees of freedom. This
may be of interest to string theory motivated work, where, in many
cases a large or infinite number of scalar fields appear in the
low energy effective 4-dimensional theory (see
\cite{Douglas:2006es} for a review).

The anti-de-Sitter (adS) solutions, while attainable, have
distinct numerical issues related with finding solutions over a
wide parameter range. However, the critical temperature generally
exhibited by adS solutions, due to the thermal bath, may result in
interesting behaviour when Hawking evaporation is considered if
the unique thermodynamic features of the $N=2$, $\Lambda = 0$
system shown below are also manifest in the $\Lambda = 1$ case.
This is left to further work although again solutions for a given
parameter range appear in \cite{leiththesis}.

\section{Thermodynamic behaviour}
\label{sec:thermodynamics}
The surface gravity\footnote{We are considering surface gravity
here, temperature may not be well-defined due to the lack of $T=0$
black holes. In the discussion we use temperature and surface
gravity interchangeably as $T \sim \kappa$ still holds.} for a
black hole in these coordinates is given by \cite{Nielsen:2005af}
\beq \kappa = \frac{A(r_{h})}{4m(r_h)}(1-2m'(r_{h})) \eeq
Clearly we will have zero-temperature black hole solutions
($\kappa = 0$) if $m'(r_{h}) = 1/2$. In the horizon expansion
given above this would correspond to $m_{1}=1/2$ while from the
equation of motion (\ref{eomGtt}) we find
\beq m_{1} = \frac{P^{2}}{2r^{2}}
\left(\frac{\lambda_{1}^{2}e^{-2g_{1}\Phi_{h}}+\lambda_{2}^{2}e^{-2g_{2}\Psi_{h}}}{\lambda_1^2
+ \lambda_2^2}\right). \eeq
Hence, $\kappa = 0$ when
\beq m(r_h)\Big|_{\kappa=0} =
\frac{P}{2}\sqrt{\frac{\lambda_{1}^{2}e^{-2g_{1}\Phi_h}+\lambda_{2}^{2}e^{-2g_{2}\Psi_h}}{\lambda_1^2
+ \lambda_2^2}}, \label{cond1} \eeq
where we have used $r_h = 2 m(r_h)$. We have denoted this limiting
case as $m(r_h)|_{\kappa = 0}$ as it turns out not to be the
extremal case. The ``extremal solution'' is the solution existing
with the maximum electromagnetic charge for a given mass and
scalar charge.

We find that the separate condition given in Mignemi and
Wiltshire~\cite{Mignemi:2004ms},
\beq P_\infty^2 \leq M_{ADM}^2 + \sum^N_{i=1} \Phi_{-1,\, i}^2,
\label{cond3} \eeq
is the constraint for extremal black holes. This constraint has no
thermodynamic significance but the equality indicates the
degenerate horizon. This is a novel situation as when
(\ref{cond1}) and (\ref{cond3}) are considered we find $P_{\infty,
extremal} < P_{\infty, \kappa = 0}$. For the degenerate horizon,
given by the equality in (\ref{cond3}), we find that the horizon
becomes singular and hence does not have a well-defined surface
gravity. This is indicated by divergence of the coefficient $m_1$
in (\ref{expansions}).

The limiting cases, however, still indicate extremal black hole
solutions with non-zero, finite surface gravity for general
couplings, $g_i$. Such results have been found previously for
specific couplings~\cite{Gibbons:1987ps}.

The limiting behaviour due to (\ref{cond3}) is shown in fig.
\ref{figure2}, where $\Phi_{-1}$ is defined by
\beq \Phi_{-1} =
g_{1}\lambda_{1}^{2}P^{2}\int^{\infty}_{r_{h}}\frac{e^{-2g_{1}\Phi}}{r^{2}}A\d
r, \eeq
and similarly for $\Psi_{-1}$~\cite{Mignemi:2004ms}.
\begin{figure*}[]
\centering
\mbox{\resizebox{8.0cm}{!}{\includegraphics{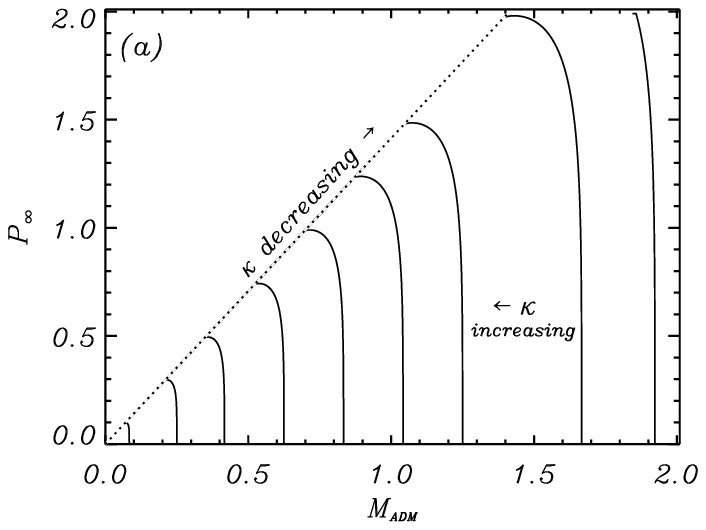}}}
\mbox{\resizebox{8.0cm}{!}{\includegraphics{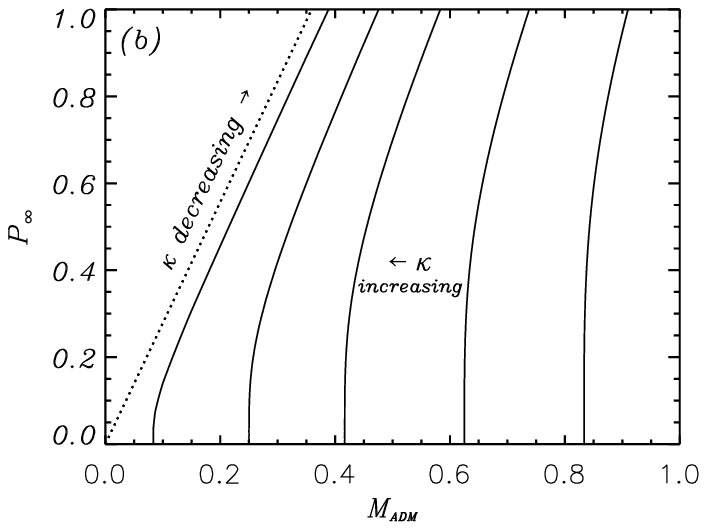}}}
\caption{Contour plots showing the behaviour of the surface
gravity; $\lambda_1=\lambda_2=1.0$ with horizon values $\Phi(r_h)
= 1.0$ and $\Psi(r_h) = 0.1$: (a) $g_1 = g_2 =1.0$; (b)
$g_1=g_2=3.0$. Note the change in behaviour from
Reissner-Nordstr\"{o}m-like solutions in (a) where the specific
heat changes sign for $P_\infty = const$ as it moves away from the
extremal limit while in (b) the specific heat for $P_\infty =
const$ is always negative and mimics the well-known Kaluza-Klein
examples. The extremal limit is shown as a dotted line as it is
not an `isotherm', instead the surface gravity decreases with
increasing $P_\infty$.} \label{figure3}
\end{figure*}

\begin{figure}[]
\epsfig{figure=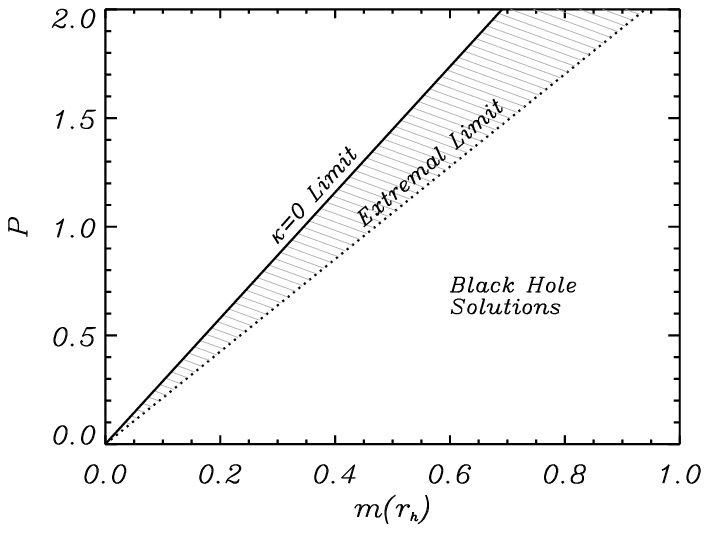, width=8.0cm, height=6.0cm} \caption{The
dotted line shows the constraint (\ref{cond3}) while the solid
line indicates the limit imposed by (\ref{cond1}) for
$\lambda_1=\lambda_2=g_1=g_2=1.0$, $\Phi(r_h)=1.0$ and $\Psi(r_h)
= 0.1$ while varying over horizon values for $m(r_h)$ and the
`bare' $P$ values.} \label{figure4}
\end{figure}

Contour plots of the surface gravity show
Reissner-Nordstr\"{o}m-like solutions when $g_1=g_2=1.0$ and
Kaluza-Klein-like solutions when $g_1 = g_2 = 3.0$ see Fig.
\ref{figure3}(a) and Fig. \ref{figure3}(b) respectively. The
`specific heat', defined as $C \equiv (\partial M_{ADM} /
\partial T ) |_{P_\infty} \sim (\partial M_{ADM} / \partial \kappa
) |_{P_\infty}$, changes sign in fig. \ref{figure3}(a) while it is
always negative for fig. \ref{figure3}(b). Unfortunately we do not
possess the computational power to find the limit of the coupling
gradients that produce these two types of solutions.

The extremal limit in these cases is not an `isotherm' but instead
tends to finite non-zero values where the surface gravity is
decreasing with increasing $P_\infty$. The contours mimic those
found in \cite{Gibbons:1987ps} but with a region excluded due to
the constraint (\ref{cond3}). Fig. \ref{figure4} shows this
graphically. However, we would caution that the extremal solution
falls into a different class from those solved by the numerical
method implemented here. As the value of $m^\prime(r_h)$ diverges
we do not have well-defined solutions and hence no surface
gravity. We therefore only comment on the limiting behaviour for
the extremal cases.

\section{Discussion}
\label{sec:discussion}

We have numerically demonstrated linearly stable black hole
solutions with contingent primary hair. The condition
(\ref{condgen}), as previously derived in \cite{Mignemi:2004ms},
gives $N+1$ asymptotic charges for $N$ scalar fields with two
being $M_{ADM}$ and $P_\infty$ (in the non-zero magnetic monopole
solution considered here) with the other $N-1$ charges being the
$1/r$ coefficients of the asymptotic expansion of $N-1$ of the
scalar fields,
\{$\Phi_{-1,1}$,$\Phi_{-1,2}$,...,$\Phi_{-1,N-1}$\}. This
violation of the no-hair theorems is, however, not entirely within
the confines of the premise under which the theorems were
originally derived as we have non-minimal coupling between the
scalar field and the $U(1)$ gauge field.

The solutions here may help to shed some light on black hole
solutions to the low energy effective 4-dimensional string theory
when coupled with further corrections, such as, higher-order
gravity terms~\cite{Chen:2006ge}, another $U(1)$ field or the
inclusion of scalar potentials. Chen {\it et al.} have found
constraints on the value of the coupling in the single scalar case
when a Gauss-Bonnet term is introduced. This appears to limit the
applicability in string theory motivated situations. However, this
constraint would possibly be weakened by additional scalar fields,
similar to the case for the slope of the potential when additional
scalar fields are considered in cosmology~\cite{Liddle:1998jc}.

The result of major interest is that the solutions are bounded by
(\ref{cond3}) and do not contain the $\kappa = 0$ case. At the
extremal limit no surface gravity is defined, it does, however,
limit to finite, non-zero values. Previously this behaviour has
only been seen in \cite{Gibbons:1987ps} when considering the
limiting case between Reissner-Nordstr\"{o}m and Kaluza-Klein type
solutions for a single scalar field with $g=\sqrt{D-1}$. Although
(\ref{condgen}) limits the number of independent asymptotic
charges, it does not allow us further insight into the nature of
the horizon. The Gibbons-Maeda solution allowed analysis of the
horizon in the extremal case, indicating a singularity. This would
seem likely in the present case as $m^\prime(r)$ diverges at the
horizon indicating a curvature singularity.

\end{document}